\documentclass[letterpaper, 10 pt, conference]{ieeeconf}  
\IEEEoverridecommandlockouts
\usepackage{balance}
\usepackage{color}
\usepackage{caption}
\usepackage{subcaption}
\usepackage{cite}
\usepackage{empheq}
\usepackage{mathrsfs}
\usepackage{enumerate}
\usepackage{siunitx}

\makeatletter
\let\proof\@undefined                        
\let\endproof\@undefined                  
\makeatother
\usepackage{graphicx,amssymb,amstext,amsmath,amsthm}

\usepackage[bookmarks=true]{hyperref}
\usepackage{algorithm,algorithmicx,algpseudocode}
\algnewcommand{\algorithmicgoto}{\textbf{go to}}%
\algnewcommand{\Goto}[1]{\algorithmicgoto~\ref{#1}}%
\algnewcommand{\LineComment}[1]{\Statex \(\triangleright\) #1}
\algnewcommand{\LineCommentN}[1]{\Statex \hspace{1cm}\(\triangleright\) #1}

\usepackage{multirow}
\usepackage{stfloats}

\usepackage{tikz}
\usepackage{circuitikz}

\makeatletter
\pgfcircdeclarebipole{}{\ctikzvalof{bipoles/interr/height 2}}{spst}{\ctikzvalof{bipoles/interr/height}}{\ctikzvalof{bipoles/interr/width}}{

    \pgfsetlinewidth{\pgfkeysvalueof{/tikz/circuitikz/bipoles/thickness}\pgfstartlinewidth}

    \pgfpathmoveto{\pgfpoint{\pgf@circ@res@left}{0pt}}
    \pgfpathlineto{\pgfpoint{.6\pgf@circ@res@right}{\pgf@circ@res@up}}
    \pgfusepath{draw}   
}

\def\pgf@circ@spst@path#1{\pgf@circ@bipole@path{spst}{#1}}
\tikzset{switch/.style = {\circuitikzbasekey, /tikz/to path=\pgf@circ@spst@path, l=#1}}
\tikzset{spst/.style = {switch = #1}}
\makeatother

\newtheorem{prop}{Proposition} 
\newtheorem{cor}{Corollary}
\newtheorem{thm}{Theorem}
	\newtheorem{assumption}{Assumption}
\newtheorem{lem}{Lemma}
\newtheorem{defn}{Definition}

\newtheorem{problem}{Problem}

\newcommand{\yong}[1]{{\color{black} #1}}
\newcommand{\yo}[1]{{\color{black} #1}}

\newcommand{\mo}[1]{{\color{black} #1}}
\begin{document}
\allowdisplaybreaks

\title{\LARGE \bf Resilient Interval Observer for Simultaneous Estimation of States, Modes and Attack Policies}

\author{%
Mohammad Khajenejad \quad\quad Zeyuan Jin \quad\quad Sze Zheng Yong \\
\thanks{
The authors are with the School for Engineering of Matter, Transport and Energy, Arizona State University, Tempe, AZ, USA (e-mail:  mkhajene@asu.edu, zjin43@asu.edu, szyong@asu.edu). This work was supported in part by NSF grant CNS-1943545.}
\vspace{-0.4cm}
}
\maketitle
\thispagestyle{empty}
\pagestyle{empty}

\begin{abstract}
 This paper considers the problem of designing interval observers for hidden mode switched nonlinear systems with bounded noise signals that are compromised by false data injection and switching attacks. The proposed observer consists of three components: i) a bank of mode-matched observers, which 
 simultaneously estimates the corresponding mode-matched continuous states and discrete states (modes), as well as learns a model of the unknown attack policy, ii) a mode observer that eliminates the incompatible modes based on a residual-based set-membership criterion, and iii) a global fusion observer that combines the outputs of i) and ii). 
Moreover, in addition to showing the correctness, stability and convergence of the mode-matched estimates, we provide sufficient conditions to guarantee that all false modes will be eliminated after sufficiently large finite time steps, i.e., the system is mode-detectable under the proposed observer.
 \end{abstract}
 
\vspace{-0.1cm}
\section{Introduction}
Computation and communication constituents are tightly intertwined in Cyber-Physical Systems (CPS). While this coupling can enhance the functionality of control systems and improve their performance, it might also become a source of vulnerability to faults or attacks.  On the other hand, given various sources of real world uncertainties, complete information/direct knowledge of the decisions and intentions of other systems/agents, is not available to autonomous decision makers, e.g., self-driving cars or robots.
These safety-critical systems can be  studied using a general framework of \emph{hidden mode hybrid/switched systems} (HMHS, see, e.g., \cite{yong2018switching} and references therein). The ability to estimate the continuous states, attacks/unknown inputs and modes/discrete states of such systems is important for monitoring them as well as for designing safe and secure (optimal) feedback controllers.
	
\emph{Literature review.} 
There has been a relatively large body of literature on the problem of designing filters/observers for hidden mode systems without considering unknown inputs/faults/data injection attacks, e.g., in \cite{Bar-Shalom.2002} and references therein. For stochastic settings, extensions were proposed, e.g., in \cite{yong2018switching}, to obtain state and unknown input \emph{point} estimates, i.e., the most likely or best single estimates. {However, especially when hard guarantees or bounds are important, it might be preferable to consider \emph{set-valued} uncertainties, e.g., bounded-norm noise. Moreover, probabilistic distributions/stochastic characteristics of uncertainty are often unavailable in real world applications. Consequently, to estimate the ``set" of compatible states, \emph{set-valued} or \emph{set-membership} observers, e.g., \cite{blanchini2012convex}, have been proposed. Later, the study in  \cite{khajenejad2020simultaneousnonlinear} extended this framework to include estimation of unknown inputs/attacks. Nonetheless, these approaches are not directly applicable to systems with hidden modes that are considered in this paper.}

A common approach to consider hidden modes for representing attack or fault models 
 is to construct \emph{residual} signals (see, e.g., \cite{yong2018switching,Bar-Shalom.2002,giraldo2018survey}),  
where to 
distinguish between 
consistent and inconsistent modes, 
some residual-based criteria/thresholds 
are used. 
The work in \cite{nakahira2018attack} presented a robust control-inspired 
\yong{approach for linear systems}
{with bounded-norm noise that} 
consists of 
local estimators, 
residual detectors, and a global fusion detector for resilient state estimation against sparse data injection attacks. 
Similar residual-based approaches have been proposed 
for uniformly observable nonlinear systems 
in \cite{kim2018detection} and some classes of nonlinear systems in \cite{chong2020secure}, where only sensors were compromised by sparse attacks, which is a special case of hidden mode switched systems discussed in our previous works \cite{yong2018switching,khajenejad2021simultaneousNAHS}. 

 On the other hand, when 
 the system model is not exactly known, in order to find \emph{a set of dynamics} that \emph{frame/bracket} the unknown system dynamics \cite{Milanese2004SetMI},
 set-valued data-driven approaches have been developed 
 to use 
input-output data to \emph{abstract} or over-approximate 
unknown dynamics or functions \cite{Milanese2004SetMI,zabinsky2003optimal},
 under the assumption that the unknown dynamics is 
 continuous, e.g., \cite{zabinsky2003optimal}.
In our previous work \cite{khajenejad2021intervalACC}, we leveraged interval observers for such data-driven models,  for resilient state and data injection attack estimation, assuming that the attack signal has an unknown dynamics. In this work, we assume mode/switching attacks in addition to data injection attacks, where the attack signals are governed by an unknown and to-be-learned \emph{attack policy}. 

 \emph{Contributions.}
  To tackle this problem, leveraging a multiple-model framework proposed in our previous works \cite{khajenejad2019simultaneousmode,khajenejad2021simultaneousNAHS}, we first design a bank of mode-matched set-valued observers, where we combine a model-based interval observer approach used in \cite{khajenejad2021intervalACC,khajenejad2020simultaneousfullCDC}, with our previously introduced set-membership learning technique \cite{Jin2020datadriven}, to derive set-valued mode-matched estimates for the states and attack signal values, as well as to learn model abstractions/over-approximations for the attack policy, where we derive several desired properties for the mode-matched estimates, such as 
  \emph{correctness}, \emph{stability} and \emph{convergence}. 
  Then, we introduce a novel \emph{elimination-based} mode observer, based on a set-membership criterion, to eliminate inconsistent modes from the bank of observers.   Furthermore, we provide sufficient conditions for \emph{mode-detectability}, i.e., 
 all false modes will be eventually ruled out under some reasonable assumptions. Finally, we illustrate the performance of our proposed approach by applying it on a power system 
 example. 

\section{Preliminaries}

\emph{Notation.} $\mathbb{R}^n$, $\mathbb{R}^{n \times m}$ and $\mathbb{D}_{n}$ denote the $n$-dimensional Euclidean space, the space of $n$ by $m$ matrices and the set of all diagonal matrices in $\mathbb{R}^{n \times n}$ with their diagonal arguments being $0$ or $1$.
For vectors $v,w \in \mathbb{R}^n$ and a matrix $M \in \mathbb{R}^{p \times q}$, $\|v\|\triangleq \sqrt{v^\top v}$ and $\|M\|$ denote their (induced) $2$-norm, {and} 
$v \leq w$ is an element-wise inequality. 
The transpose, 
Moore-Penrose pseudoinverse{, $(i,j)$-th element} 
and rank of $M$ are given by $M^\top$, 
$M^\dagger${, $M_{i,j}$} 
and ${\rm rk}(M)$, while $M_{(r:s)}$ is a sub-matrix of $M$, consisting of its $r$-th through $s$-th rows, and 
its row support is  $r=\textstyle{\mathrm{rowsupp}}(M) \in \mathbb{R}^p$, where $r_i=0$ if the $i$-th row of $M$ is zero and $r_i=1$ otherwise, $\forall i \in \{1\dots p \}$. Also, $M^+ \triangleq \max(M,0_{p \times q}),M^{-} \triangleq M^+-M$ and 
$|M| \triangleq M^++M^{-}$. $M$ is a non-negative matrix, if $M_{i,j} \geq 0, \forall (i,j) \in \{1\dots p\} \times \{1 \dots q \}$. 

Next, we introduce some useful definitions and results.
\begin{defn}[Interval, Maximal and Minimal Elements, Interval Width]\label{defn:interval}
{An (multi-dimensional) interval {$\mathcal{I}  \subset 
\mathbb{R}^n$} is the set of all real vectors $x \in \mathbb{R}^n$ that satisfies $\underline{s} \le x \le \overline{s}$, where $\underline{s}$, $\overline{s}$ and $\|\overline{s}-\underline{s}\|$ are called minimal vector, maximal vector and width of $\mathcal{I}$, respectively}.
\end{defn}
\begin{prop}[{Slight Generalization of \cite[Lemma 2]{efimov2012interval}}]\label{prop:uncertain_bounding}
Let $B \in \mathbb{IR}^{n \times p}$ be an interval matrix satisfying $\underline{B} \leq B \leq \overline{B}$. 
\begin{enumerate}[i)]
\item if $A \in \mathbb{R}^{m \times n}$ is a constant matrix, then $A^+\underline{B}-A^+\overline{B} \leq AB \leq A^+\overline{B}-A^+\underline{B}$. 
\item if $A \in \mathbb{IR}^{m \times n}$ is an interval matrix satisfying $\underline{A} \leq    A \leq \overline{A}$, then $\underline{A}^+\underline{B}^+-\overline{A}^+\underline{B}^--\underline{A}^-\overline{B}^++\overline{A}^-\overline{B}^- \leq AB \leq \overline{A}^+\overline{B}^+-\underline{A}^+\overline{B}^--\overline{A}^-\underline{B}^++\underline{A}^-\underline{B}^-$
\end{enumerate}
\end{prop}
\begin{proof}
The results follow from defining $x_i$ as the $i$th column of $B$, applying \cite[Lemma 2]{efimov2012interval} on $A$ and $x_i$ for all $i \in \mathbb{N}_p$ and then stacking the resulting inequalities.
\end{proof}
\begin{prop}[Parallel Affine Abstractions \cite{khajenejad2021intervalACC}]\label{prop:affine abstractions}
Let the entire space be defined as $\mathbb{X}$ and suppose that $\mathbb{X}$ is bounded. Consider the vector fields $\overline{\psi}(.),\underline{\psi}(.):\mathbb{X} \subset \mathbb{R}^{n'} \to \mathbb{R}^{m'}$ satisfying $\underline{\psi}(x) \leq \overline{\psi}(x), \forall x \in \mathbb{X}$, a (given) global parallel affine abstraction with known $(\mathbb{A}^{\psi},\overline{e}^{\psi},\underline{e}^{\psi})$  on $\mathbb{X}$, i.e., 
\begin{align} \label{eq:global_abs}
\mathbb{A}^{\psi}{x}+\underline{e}^{\psi} \leq \underline{\psi}(x) \leq \overline{\psi}(x) \leq  \mathbb{A}^{\psi} {x}+\overline{e}^{\psi}, \forall  x \in \mathbb{X}.
\end{align}  
and the 
following Linear Program (LP): 
\begin{subequations}  
\begin{align} 
\label{eq:abstraction} \hspace{-0.5cm}&\min\limits_{\theta^{\psi}_\mathcal{B},{A}^{\psi}_\mathcal{B},\overline{e}^{\psi}_\mathcal{B},\underline{e}^{\psi}_\mathcal{B}} {\theta^{\psi}_\mathcal{B}} \\
\nonumber   \hspace{-0.5cm} &s.t. ~  {A}^{\psi}_\mathcal{B} {x}_{s} \hspace{-0.05cm}+\hspace{-0.05cm}\underline{e}^{\psi}_\mathcal{B} \hspace{-0.05cm}+\hspace{-0.05cm} \sigma^{\psi} \leq \underline{\psi}({x}_{s}) \leq  \overline{\psi}({x}_{s}) \leq {A}^{\psi}_\mathcal{B} {x}_{s}\hspace{-0.05cm}+\hspace{-0.05cm}\overline{e} ^{\psi}_\mathcal{B}\hspace{-0.05cm}-\hspace{-0.05cm}\sigma^{\psi}, \\
\nonumber &\quad   \overline{e}_{\mathcal{B}}^{\psi}-\underline{e}_{\mathcal{B}}^{\psi}-2\sigma^{\psi} \leq \theta^{\psi} \mathbf{1}_{m'} , \\
 & \quad   \underline{e}^{\psi}-\underline{e}^{\psi}_\mathcal{B} \leq ({A}^{\psi}_\mathcal{B}-\mathbb{A}^{\psi}) x_{s} \leq \overline{e}^{\psi}-\overline{e}^{\psi}_\mathcal{B}, \forall  x_s \in \mathcal{V}_{\mathcal{B}}, \label{eq:guarantee}
\end{align}  
\end{subequations}
where $\mathcal{B}=[\underline{x},\overline{x}] \subseteq \mathbb{X}$ is a local interval domain 
and $\mathcal{V}_{\mathcal{B}}$ being its maximal, minimal and set of vertices, respectively, $\mathbf{1}_{m} \in \mathbb{R}^m$ is a vector of ones, $\sigma^{\psi}$ is given in \cite[Proposition 1 and (8)]{singh2018mesh} for different classes of continuous vector fields.  
Then, $({A}^{\psi}_\mathcal{B},\overline{e}^{\psi}_\mathcal{B},\underline{e}^{\psi}_\mathcal{B})$ are the local parallel affine abstraction matrices for the pair of functions $\overline{\psi}(.),\underline{\psi}(.)$ on $\mathcal{B}$, i.e., 
\begin{align} \label{eq:global_abs}
{A}^{\psi}_\mathcal{B}{x}+\underline{e}^{\psi}_\mathcal{B} \leq \underline{\psi}(x) \leq \overline{\psi}(x) \leq  {A}^{\psi}_\mathcal{B} {x}+\overline{e}^{\psi}_\mathcal{B}, \forall  x \in \mathcal{B}.
\end{align} 
\end{prop}
\begin{defn}[Mixed-Monotone Mappings and Decomposition Functions]\cite[Definition 4]{yang2019sufficient}\label{defn:mixed-monotone}
A mapping $f:\mathcal{X} \subseteq \mathbb{R}^n \rightarrow \mathcal{T} \subseteq \mathbb{R}^m$ is mixed-monotone if there exists a decomposition function $f_d:\mathcal{X} \times \mathcal{X} \rightarrow \mathcal{T}$ satisfying: 
i) $f_d(x,x)=f(x)$, ii) $x_1 \geq x_2 \Rightarrow f_d(x_1,y)\geq f_d(x_2,y)$ and iii) $y_1 \geq y_2 \Rightarrow f_d(x,y_1) \leq f_d(x,y_2)$.  
\end{defn}
\begin{prop}\cite[Theorem 1]{coogan2015efficient}\label{prop:embedding}
Let $f:\mathcal{X} \subseteq \mathbb{R}^n \rightarrow \mathcal{T} \subseteq \mathbb{R}^m$ be a mixed-monotone mapping with decomposition function $f_d:\mathcal{X} \times \mathcal{X} \rightarrow \mathcal{T}$ and $\underline{x} \leq x \leq \overline{x}$, where $\underline{x},x,\overline{x} \in \mathcal{X}$. Then $f_d(\underline{x},\overline{x}) \leq f(x) \leq f_d(\overline{x},\underline{x})$.
\end{prop}
\begin{cor}[Nonlinear Bounding]\label{cor:bounding}
Let $f:\mathcal{X} \subseteq \mathbb{R}^n \rightarrow \mathcal{T} \subseteq \mathbb{R}^m$ satisfies the assumptions in Propositions \ref{prop:affine abstractions} and \ref{prop:embedding}. Then, for all $\underline{x},x,\overline{x} \in \mathcal{X}$ satisfying $\underline{x} \leq x \leq \overline{x}$, the following inequality holds: $\underline{f} \leq f(x)  \leq \overline{f}$, where 
\begin{align}\label{eq:nonl_bounding}
\begin{array}{rl}
\overline{f}&=\min(f_d(\overline{x},\underline{x}),A^{f+}\overline{x}-A^{f-}\underline{x}+\overline{e}^f),\\
\underline{f}&=\max(f_d(\underline{x},\overline{x}),A^{f+}\underline{x}-A^{f-}\overline{x}+\underline{e}^f),
\end{array}
\end{align}
$f_d$ is a decomposition function of $f$ (cf. Definition \ref{defn:mixed-monotone}) and $A^f,\overline{e}^f,\underline{e}^f$ are the affine abstraction slope and errors of $f$, computed over the interval $[\underline{x},\overline{v}]$, through Proposition \ref{prop:affine abstractions}. 
 \end{cor}
 \begin{proof}
 The results directly follow from Propositions \ref{prop:uncertain_bounding}--\ref{prop:embedding}.
 \end{proof}
Note that the decomposition function of a 
vector field 
{is not unique and a specific one is given} 
in \cite[Theorem 2]{yang2019sufficient}{: If} a vector field $q=\begin{bmatrix} q^\top_1 & \dots & q^\top_n \end{bmatrix}^\top:X \subseteq \mathbb{R}^n \rightarrow \mathbb{R}^m$ is differentiable and its partial derivatives are bounded with known bounds, i.e., $\frac{\partial q_i}{\partial x_j} \in (a^q_{i,j},b^q_{i,j}), \forall x \in X \in \mathbb{R}^n$, where $a^q_{i,j},b^q_{i,j} \in {{\mathbb{R}}}$, then $q$ is mixed-monotone with a decomposition function $q_d=\begin{bmatrix} q^\top_{d1} & \dots & q^\top_{di} & \dots q^\top_{dn} \end{bmatrix}^\top$, where $q_{di}(x,y)=q_{i}(z)+(\alpha^q_i-\beta^q_i)^\top (x-y), \forall i \in \{1,\dots,n\}$, and $z,\alpha^q_i,\beta^q_i \in \mathbb{R}^n$ can be computed in terms of $x, y, a^q_{i,j}, b^q_{i,j}$ as given in \cite[(10)--(13)]{yang2019sufficient}. Consequently, for $x=[x_1\,\dots \, x_j \, \dots\, x_n]^\top$, $y=[y_1\,\dots\, y_j\, \dots\, y_n]^\top$, we have 
\begin{align}
q_{d}(x,y)=q(z)+C^q(x-y),  \label{eq:decompconstruct} 
\end{align}
where $C^q \hspace{-.1cm}  \triangleq \hspace{-.1cm}  \begin{bmatrix} [\alpha^q_1-\beta^q_1] & \hspace{-0.2cm}\dots \hspace{-0.2cm} & [\alpha^q_i-\beta^q_i] & \dots [\alpha^q_m-\beta^q_m] \end{bmatrix}^\top \hspace{-.2cm} \in \hspace{-.1cm}  \mathbb{R}^{m \times n}$, 
with 
$\alpha^q_i,\beta^q_i$ given in \cite[(10)--(13)]{yang2019sufficient}, $z=[z_1 \dots z_j \dots z_m]^\top$ and $z_j=x_j$ or $y_j$ (dependent on the case, cf. \cite[Theorem 1 and (10)--(13)]{yang2019sufficient} for details). On the other hand, when the precise lower and upper bounds, $a_{i,j}, b_{i,j}$, of the partial derivatives are not known or are hard to compute, we can obtain upper and lower approximations of the bounds by using Proposition \ref{prop:affine abstractions} 
with 
the slopes \yong{set} to zero, or by leveraging interval arithmetics \cite{jaulin2002nonlinear}. 

\section{Problem Statement} \label{sec:Problem}
\noindent\textbf{\emph{System Assumptions.}} 
Consider a 
discrete-time hidden mode switched nonlinear system with bounded-norm noise and unknown inputs (i.e., 
 a hybrid system with nonlinear and noisy system dynamics in each mode, and 
 the mode and some inputs are not known/measured): 
\begin{align} \label{eq:sys_desc}
\begin{array}{rll}
\hspace{-0.25cm}
 x_{k+1}\hspace{-0.15cm}&=\hat{f}^q (x_k,u^{q}_k,G^{q} d^{q}_{k},w_k)\triangleq f^q(x_k,d^q_k,w_k), 
\\ 
y_k&=\hat{g}^q( x_k,u^{q}_k,H^{q} d^{q}_k,v_{k})\triangleq g^q(x_k,d^q_k,v_k), \\
d^q_k&=\hat{\mu}^q(x_k,u^q_k) \triangleq \mu^q(x_k), 
\end{array}
\end{align}
where $x_k \in \mathbb{R}^n$ is the continuous system state and $q \in \mathcal{Q}=\{1,2,\dots,Q\}$ is the hidden discrete state or \emph{mode}. For each (fixed) mode $q$, 
$u^q_k \in U^q_{k} \subset \mathbb{R}^m$ is the \emph{known} input, $d^q_k \in \mathbb{R}^p$ is the unknown but \emph{sparse} input, i.e., every vector $d^q_k$ has precisely $\rho \in \mathbb{N}$ nonzero elements where $\rho$ is a known parameter and $y_k \in \mathbb{R}^l$ is the measured output. The unknown input signal $d^q_k$ is considered as the realization of an attacker's \emph{unknown policy} $\mu^q: \mathbb{R}^n \times \mathbb{R}^m \times \mathbb{R}^s \to \mathbb{R}^p$, which is an unknown mapping from state and known input to the set of attack signals. 
Moreover, $w_k \in \mathcal{W} \triangleq [\underline{w},\overline{w}] \subset \mathbb{R}^{n_w}$ and $v_k \in \mathcal{V} \triangleq [\underline{v},\overline{v}] \subset \mathbb{R}^{n_v}$ are bounded process and measurement disturbances with known minimal and maximal values $\underline{w},\overline{w},\underline{v},\overline{v}$, respectively.
Further,  the mappings $f,g$, as well as the matrices $G^q \in \mathbb{R}^{n \times p}$ and $H^q \in \mathbb{R}^{l \times p}$ are known.
 
 More precisely, $G^q$ and $H^q$ represent the different hypothesis for each mode $q \in \mathcal{Q}$, about the sparsity pattern of the unknown inputs, which in the context of sparse attacks corresponds to which actuators and sensors are attacked or not attacked. In other words, we assume that $G^q=G\mathbb{I}^q_G$ and $H^q=H\mathbb{I}^q_H$ for some input matrices $G \in \mathbb{R}^{n \times t_a} $ and $H \in \mathbb{R}^{l \times t_s} $, where 
$t_a$ and $t_s$ are the number of vulnerable actuator and sensor signals respectively. Note that $\rho^q_a \leq t_a \leq m$ and  $\rho^q_s \leq t_s \leq l$, where $\rho^q_a$ ($\rho^q_s$) is the number of attacked actuator (sensor) signals and clearly cannot exceed the number of vulnerable actuator (sensor) signals, which in turn cannot exceed the total number of actuators (sensors). Further, we assume that the maximum number of unknown inputs/attacks in each mode is known and equals $\rho=\rho_a+\rho_s$ (sparsity assumption). Moreover, {the \emph{index matrix}} $\mathbb{I}^q_G \in \mathbb{R}^{t_a \times \rho}$ ($\mathbb{I}^q_H \in \mathbb{R}^{t_s \times \rho}$) represents the sub-vector of $d_k \in \mathbb{R}^{\rho}$ \yong{that indicates} signal magnitude attacks on the actuators (sensors). 

We are interested in estimating the state trajectories, as well as the unknown mode and the attack policy mapping in the system in \eqref{eq:sys_desc}, when they are initialized in a given interval
$\mathcal{X}_0 \subset \mathcal{X} \subset \mathbb{R}^n$.
Furthermore, we assume the following:
\begin{assumption}\label{ass:mixed_monotone}
The vector fields $f,g$ are known, Lipschitz continuous and mixed-monotone. Moreover, the values of the input $u^q_k$ and output/measurement $y_k$ signals are known at all times and for all modes. The set of all possible modes, $\mathcal{Q}$, is also known.
\end{assumption}
\begin{assumption}\label{ass:unknown_Lip}
Given mode $q$, the attacker's policy mapping $\mu^q(\cdot)=[\mu^{q\top}_1(\cdot),\dots,\mu^{q\top}_p(\cdot)]^\top$ is unknown, but each $\mu^q_j(\cdot),\forall j \in \{1,\dots,p\}$ is known to be Lipschitz continuous. Moreover, for simplicity and without loss of generality we assume that the Lipschitz constants $L^{\mu^q}_j, \forall j \in \{1,\dots,p\}$ are known, otherwise, they can be estimated 
with any desired precision 
using the approach in \cite[Equation (12) and Proposition 3]{Jin2020datadriven}. 
\end{assumption}  
\begin{assumption}\label{ass:fixed_true_mode}
There is only one ``true" mode, i.e. the true mode $q^* \in \mathcal{Q}$ is constant over time.
\end{assumption}
 
 Note that the approach in our paper can be easily extended to handle mode-dependent $f$, $g$, $\overline{w},\underline{w}$, $\overline{v}$ and $\underline{v}$, but is omitted to simplify the notations. 
Further, we formally define the notions of \emph{framers}, \emph{correctness} and \emph{stability} that are used throughout the paper. 
\begin{defn}[Framers and Correct Interval Observers]\label{defn:framers}
Given a hidden mode switched nonlinear system \eqref{eq:sys_desc}, let us define the augmented state $z_k \triangleq [x^\top_k \ d^\top_k]^\top$, for all $k \in {\mathbb{K}}\triangleq \mathbb{N}\cup \{0\}$, where $d_k \triangleq d^{q^*}_k$ is the true attack signal. 
The sequences $\{\overline{z}_k,\underline{z}_k\}_{k=0}^{\infty}$ are called upper and lower framers for the augmented states of system \eqref{eq:sys_desc}, if 
$\forall k \in \yong{\mathbb{K}}, \ \underline{z}_k \leq z_k \leq \overline{z}_k$.
In other words, starting from the initial interval $z_0 \in [\underline{z}_0,\overline{z}_0]$, the true augmented state of the system in \eqref{eq:sys_desc}, $z_k$, is guaranteed to evolve within the interval flow-pipe $[\underline{z}_k,\overline{z}_k]$, for all $k \in \yong{\mathbb{K}}$. Finally, 
any algorithm that returns framers for the states of system \eqref{eq:sys_desc} is called a \emph{correct} interval observer for system \eqref{eq:sys_desc}. 
\end{defn}
\begin{defn}[Stability] \label{defn:stability}
The mode-matched observer  \eqref{eq:propagation}--\eqref{eq:ml} is stable, if the sequence of interval widths  $\{\|\Delta^{z^q}_{k}\| \triangleq \|\overline{z}^q_{k}-\underline{z}^q_{k}\|\}_{k=0}^{\infty}$ is {uniformly bounded}, and consequently, the sequence of estimation errors $\{\|\tilde{z}^q_{k}\| \triangleq \max (\|z^q_{k}-\underline{z}^q_{k}\|,\|\overline{z}^q_{k}-{z}^q_{k}\|)$ is also uniformly bounded.
\end{defn}
Using the modeling framework above, the simultaneous state, hidden mode and policy estimation problem 
is threefold and can be stated as follows:
\begin{problem}
Given a discrete-time bounded-error hidden mode switched nonlinear system with unknown inputs \eqref{eq:sys_desc} and assuming that Assumptions \ref{ass:mixed_monotone}--\ref{ass:fixed_true_mode} hold, 
\begin{enumerate}[i)]
\item Design a bank of mode-matched observers that for each mode, \yo{conditioned on the mode being the true mode,} 
 finds uniformly bounded set estimates  
of compatible (augmented) states and learns a guaranteed model abstraction of the attacker's policy. 
\item Develop a mode observer via elimination and the corresponding criteria to eliminate false modes. 
\item Find sufficient conditions for eliminating all false modes.
\end{enumerate}
\end{problem}
 
\section{Proposed Observer Design}
Leveraging a multiple-model approach similar to \cite{khajenejad2019simultaneousmode,khajenejad2021simultaneousNAHS}, for simultaneous mode, state and attack policy (SMSP) estimation, our goal in this section is to propose an observer to find set estimates $\hat{\mathcal{X}}_k$, $\hat{\mathcal{D}}_k$ and $\hat{\mathcal{Q}}_k$ for the states $x_k$, attacks $d_k$ and modes $q \in \mathcal{Q}$ at time step $k$, respectively, as well as to compute a model abstraction $\{\overline{\mu}_k,\underline{\mu}_k\}_{k \in \mathbb{K}}$ for the attack policy, such that $\underline{\mu}_k(x_k) \leq \mu(x) \leq \overline{\mu}(x_k)$ for all $k \in \mathbb{K}$.
\subsection{Multiple-Model Approach: An Overview} \label{sec:overview}
 Similar to the approach in \cite{khajenejad2019simultaneousmode}, we propose a three-step multiple-model design consisting of: (i) a bank of mode-matched interval observers to obtain mode-matched state and attack estimates, as well as mode-matched policy abstractions/over-approximations, (ii) a mode estimation algorithm to eliminate incompatible modes using residual detectors, and (iii) a global fusion observer that outputs the desired set-valued mode, attack (policy) and state estimates.
\subsubsection{Mode-Matched Set-Valued State and Attack Policy Observer} \label{sec:ULISE}
First, we design a bank of mode-matched observers, which consists of $Q \triangleq | \mathcal{Q}|$ simultaneous state, attack and policy mode-matched interval observers, designed in a similar manner as our approach in \cite{khajenejad2021intervalACC}, with the difference that in \cite{khajenejad2021intervalACC}, 
the unknown input (i.e., attack) signal \yo{is treated as a state with} 
unknown and to-be-learned dynamics, 
whereas in the current work, the attack signal is governed by an unknown policy\yo{/state feedback law}, i.e., an unknown 
function of the actual state, that should be learned/approximated. With that in mind, 
given mode $q$, each mode-mathced interval observer 
at time step $k \in \mathbb{N}$, returns 
\begin{align}
\begin{array}{rl}
\hat{\mathcal{X}}^q_k &\triangleq [\underline{x}^q_k,\overline{x}^q_k],\ \hat{\mathcal{D}}^q_k \triangleq [\underline{d}^q_k,\overline{d}^q_k], \ \{\underline{\mu}^q_k,\overline{\mu}^q_k\}, 
\end{array}
\end{align}
such that $x_k \in \hat{\mathcal{X}}^q_k$, $d_k \in \hat{\mathcal{D}}^q_k$ and  $\mu(x_k) \in [\underline{\mu}^q_k(x_k),\overline{\mu}^q_k(x_k)]$  
through the following steps (with the augmented state $z^q_k \triangleq \begin{bmatrix} x^\top_k & d^{q\top}_k \end{bmatrix}^\top$ and known $\underline{x}_{0}$ and $\overline{x}_0$ such that $\underline{x}_{0} \leq x_0 \leq \overline{x}_0$): 

\vspace{-0.1cm}
\noindent \textbf{\emph{State Propagation}}: 
\begin{subequations}
\begin{align}
\label{eq:propagation}  &\hspace{-.1cm}\begin{bmatrix} \overline{x}^{q,p}_{k} \\ \underline{x}^{q,p}_{k} \end{bmatrix}  \hspace{-.15cm}=\hspace{-.15cm} \begin{bmatrix} \min(\overline{f}_d(\overline{z}^q_{k-1},\overline{w},\underline{z}^q_{k-1},\underline{w}),\overline{x}^{q,\hat{p}}_{k} ) \\ \max(\underline{f}_d(\underline{z}^q_{k-1},\underline{w},\overline{z}^q_{k-1},\overline{w}),\underline{x}^{q,\hat{p}}_{k})\hspace{-.05cm} \end{bmatrix}\hspace{-.1cm},\hspace{-.05cm}
\end{align}
\end{subequations}
\noindent \textbf{\emph{Attack Policy Learning}}: \vspace{-0.05cm}
\begin{subequations}
\begin{align}
& \overline{\mu}^q_{k,j}(x_k) \hspace{-0.1cm}=\hspace{-0.25cm}  \min\limits_{t \in \{0,\hdots, T-1\}}  (\overline{d}^q_{k-t,j}\hspace{-0.1cm}+\hspace{-0.1cm}L^{\mu^q}_j\|x_k\hspace{-0.1cm}-\hspace{-0.1cm}\tilde{x}^q_{k-t}\|)\hspace{-0.1cm}+\hspace{-0.1cm}\varepsilon^{q,j}_{k-t} , \label{upper_func}\\
& \underline{\mu}^q_{k,j}(x_k) \hspace{-0.1cm}=\hspace{-0.25cm}  \max_{\substack{t \in \{0,\hdots, T-1\}}}  (\underline{d}^q_{k-t,j}\hspace{-0.1cm}-\hspace{-0.1cm}L^{\mu^q}_j\|x_k\hspace{-0.1cm}-\hspace{-0.1cm}\tilde{x}^q_{k-t}\|)\hspace{-0.1cm}+\hspace{-0.1cm}\varepsilon^{q,j}_{k-t} , \label{lower_func}
 \end{align}
 \end{subequations}
 \noindent \textbf{\emph{Unknown Input Estimation}}: \vspace{-0.05cm}
 \begin{subequations}
 \begin{align}
  &\begin{bmatrix}\overline{d}^{q,p}_k \\ \underline{d}^{q,p}_k \end{bmatrix}=\hspace{-0.1cm} \begin{bmatrix} {A}^{{\mu}^q+}_{k} & -{A}^{{\mu}^q-}_{k} \\ -{A}^{{\mu}^q-}_{k} & {A}^{{\mu}^q+}_{k} \end{bmatrix} \begin{bmatrix} \overline{x}^{q,p}_k \\ \underline{x}^{q,p}_k \end{bmatrix},\\
 & \overline{z}^{q,p}_{k}  \hspace{-.1cm}=\hspace{-.1cm}\begin{bmatrix} \overline{x}^{q,p^\top}_{k} & \overline{d}^{q,p^\top}_{k}  \end{bmatrix}^\top,      \underline{z}^{q,p}_{k}=\begin{bmatrix} \underline{x}^{q,p^\top}_{k} & \underline{d}^{q,p^\top}_{k}  \end{bmatrix}^\top, \label{eq:x_prop_up}
 \end{align}
 \end{subequations}
 \noindent \textbf{\emph{Measurement Update}}: \vspace{-0.05cm}
\begin{subequations}
\begin{align}
&\hspace{-2cm} \begin{bmatrix} \overline{z}^q_{k} &\underline{z}^q_{k} \end{bmatrix}= \lim_{i \to \infty} \begin{bmatrix} \overline{z}^{q,u}_{i,k} & \underline{z}^{q,u}_{i,k} \end{bmatrix}, \label{eq:mup} \\
  &\hspace{-2cm}  \begin{bmatrix} \overline{x}^q_k & \underline{x}^q_k \\ \overline{d}^q_k &  \underline{d}^q_k \end{bmatrix}=\begin{bmatrix} \overline{z}^q_{k,(1:n)} & \underline{z}^q_{k,(1:n)} \\ \overline{z}^q_{k,(n+1:n+p)} & \underline{z}^q_{k,(n+1:n+p)} \end{bmatrix}, \label{eq:ml} 
 \end{align}
 \end{subequations}
 
 \vspace{-0.15cm}

\noindent with $j \in \{1\dots p\}$, where $\{\tilde{x}^q_{k-t}=\frac{1}{2}(\overline{x}^q_{k-t}+\underline{x}^q_{k-t})\}_{t=0}^{k}$ and $\{\overline{d}^q_{k-t},\underline{d}^q_{k-t}\}_{t=0}^{k}$ are the \emph{augmented} input-output data set. At each time step $k$, the augmented data set constructed from the estimated framers gathered from the initial to the current time step, is used in the model learning step to recursively derive over-approximations of the unknown function $\mu^q(\cdot)$, i.e., $\{\overline{\mu}^q_k(.),\underline{\mu}^q_k(.)\}$ 
 by applying \cite[Theorem 1]{Jin2020datadriven}. 
 In addition, 
\begin{align}
  \begin{bmatrix} \overline{x}^{q,\hat{p}}_{k} \\ \underline{x}^{q,\hat{p}}_{k} \end{bmatrix} \ =\mathbb{A}^{q,f}_k\begin{bmatrix} \overline{z}^{q}_{k-1} \\ \underline{z}^q_{k-1} \end{bmatrix}\hspace{-.1cm}+\mathbb{W}^{q,f}_k \begin{bmatrix} \overline{w} \\ \underline{w}\end{bmatrix}+\begin{bmatrix} \overline{e}^{q,f}_{k} \\ \underline{e}^{q,f}_{k} \end{bmatrix}, \label{eq:x_prop_abst} 
\end{align}
with  
$\mathbb{J}^{q,s}_k\hspace{-.1cm}=\hspace{-.1cm}\begin{bmatrix} J^{q,s+}_{k} & -J^{q,s-}_{k} \\ -J^{q,s-}_{k} &J^{q,s+}_{k} \end{bmatrix}$, $\varepsilon^{q,j}_{k-t}\hspace{-.1cm}=\hspace{-.1cm}2L^{\mu^q}_j\|\overline{x}^q_{k-t}-\underline{x}^q_{k-t}\|,$ $\forall \mathbb{J} \in \{\mathbb{A},\mathbb{W}\}, s \in \{f,\mu\}, J \in \{A,W\}$. 
Moreover, the \emph{sequences of updated framers} $\{\overline{z}^{q,u}_{i,k},\underline{z}^{q,u}_{i,k}\}_{i=1}^{\infty}$ are iteratively computed  as follows:  
\begin{align}
\label{eq:zupp}&\begin{bmatrix} \overline{z}^{q,u}_{0,k} & \underline{z}^{q,u}_{0,k} \end{bmatrix} = \begin{bmatrix} \overline{z}^{q,p}_k & \underline{z}^{q,p}_k \end{bmatrix},  \quad  \forall i \in \{1\dots \infty \}: \\
&\begin{bmatrix}\overline{z}^{q,u}_{i,k} \\ \underline{z}^{q,u}_{i,k} \end{bmatrix}\hspace{-.1cm}=\hspace{-.1cm}\begin{bmatrix} \min(A^{q,g\dagger +}_{i,k} \overline{\alpha}^q_{i,k}\hspace{-.1cm}-\hspace{-.1cm}A^{q,g\dagger -}_{i,k} \underline{\alpha}^q_{i,k}\hspace{-.1cm}+\hspace{-.1cm}\omega^q_{i,k},\overline{z}^{q,u}_{i-1,k}) \\ \max(A^{q,g\dagger +}_{i,k} \underline{\alpha}^q_{i,k}\hspace{-.1cm}-\hspace{-.1cm}A^{q,g\dagger -}_{i,k} \overline{\alpha}^q_{i,k}\hspace{-.1cm}-\hspace{-.1cm}\omega^q_{i,k},\underline{z}^{q,u}_{i-1,k}) \end{bmatrix}\hspace{-.1cm},\hspace{-.2cm} \label{eq:iter_update}
\end{align} 

\vspace{-.5cm}
\noindent where
\begin{align}
&\begin{bmatrix}\overline{\alpha}^q_{i,k} \\ \underline{\alpha}^q_{i,k} \end{bmatrix}=\begin{bmatrix} \min(\overline{t}^q_{i,k},A^{q,g+}_{i,k}\overline{z}^{q,u}_{i-1,k}-A^{q,g-}_{i,k}\underline{z}^{q,u}_{i-1,k}) \\ \max(\underline{t}^q_{i,k},A^{q,g+}_{i,k}\underline{z}^{q,u}_{i-1,k}-A^{q,g-}_{i,k}\overline{z}^{q,u}_{i-1,k}) \end{bmatrix}, \label{eq:alpha}\\
&\begin{bmatrix} \overline{t}^q_{i,k} \\ \underline{t}^q_{i,k} \end{bmatrix}\hspace{-.15cm}=\hspace{-.15cm}\begin{bmatrix} y_k \\ y_k \end{bmatrix} \hspace{-.15cm}+\hspace{-.15cm}\begin{bmatrix} W^{q,g-}_{i,k} & -W^{q,g+}_{i,k} \\ \ -W^{q,g+}_{i,k} & W^{q,g-}_{i,k} \end{bmatrix}\hspace{-.15cm}  \begin{bmatrix}\overline{v} \\ \underline{v}\end{bmatrix}\hspace{-.15cm}-\hspace{-.15cm}\begin{bmatrix}\underline{e}^{q,g}_{i,k} \\ \overline{e}^{q,g}_{i,k} \end{bmatrix}\hspace{-.05cm},\hspace{-.05cm} \label{eq:t} 
\end{align}
and $\omega^q_{i,k}\hspace{-.1cm}=\hspace{-.1cm}\kappa \textstyle{\mathrm{rowsupp}}(I-A^{q,g\dagger}_{i,k}A^{q,g}_{i,k}),$ $\forall i \in \{1\dots \infty\}$.
%
In addition, $(A^{q,s}_{k},W^{q,s}_{k},\overline{e}^{q,s}_{k},\underline{e}^{q,s}_{k})$ for $s \in \{f,\mu\}$
 and $(A^{q,g}_{i,k},W^{q,g}_{i,k},\overline{e}^{q,g}_{i,k},\underline{e}^{q,g}_{i,k})$ 
are solutions to the problem \eqref{eq:abstraction} 
for the corresponding functions $\{\underline{g}^q(\cdot)=\overline{g}^q(\cdot)=g^q(\cdot)\}$, $\{\underline{f}^q(\cdot)=\overline{f}^q(\cdot)=f^q(\cdot)\}$ and $\{\overline{\mu}^q_k(\cdot),\underline{\mu}^q_k(\cdot)\}$, on the intervals $[\begin{bmatrix} \underline{z}^{q,u\top}_{i-1,k} & \underline{v}^\top  \end{bmatrix}^\top, \begin{bmatrix} \overline{z}^{q,u\top}_{i-1,k} & \overline{v}^\top \end{bmatrix}^\top]$ for $g^q$, $[\begin{bmatrix} \underline{z}_{k-1}^{q\top}  \underline{w}^\top  \end{bmatrix}, \begin{bmatrix} \overline{z}_{k-1}^{q\top} & \overline{w}^\top \end{bmatrix}^\top]$ for $f^q$ and $\begin{bmatrix} \underline{x}_{k}^{q,p\top} & \overline{x}_{k}^{q,p\top}  \end{bmatrix}^\top]$ for $\overline{\mu}^q_k$, $\underline{\mu}^q_k$, respectively, at time $k$ and iteration $i$, while 
 $\kappa$ is a very large positive real number (infinity) and
 $\overline{f}^q_{d},\underline{f}^q_d$ 
  are the bounding function based on \eqref{eq:decompconstruct}.  
\subsubsection{Mode Estimation Observer}
To estimate the set of compatible modes, we consider a membership-based elimination approach that checks if residual signals are within some compatible intervals. We first define the mode-matched residual signal $r^{q}_k$ as follows. 
\begin{defn} [Residuals] \label{defn:computedresidual}
For each mode $q$ at time step $k$, 
 the residual signal $r^q_k$ is defined as:
\begin{align}\label{eq:res_def}
r^{q}_k \triangleq y_k-\frac{1}{2}(\overline{g}^q_k+\underline{g}^q_k),
\end{align}
where $\overline{g}^q,\underline{g}^q$ are the bounding signals based on \eqref{eq:nonl_bounding} applied on the mapping $g^q(\cdot)$. 
\end{defn}
Then, we eliminate a specific mode $q$, if its corresponding residual signal $r^q_k$ violates to be within an interval given in the following proposition \ref{prop:res_membership}.
\begin{prop}[Mode Elimination Criterion]\label{prop:res_membership}
Mode $q$ is not a true mode if 
\begin{align}\label{eq:mode_elimination}
r^q_k \notin \mathcal{R}^q_k \triangleq  \frac{1}{2} [-(\overline{g}^q_k-\underline{g}^q_k), \overline{g}^q_k-\underline{g}^q_k] 
\end{align}
\end{prop}
\begin{proof}
If $q$ is the true mode, then $y_k=g^q(x_k,d_k,v_k)$ by \eqref{eq:sys_desc}. Consequently, $y_k \in [\underline{g}_k,\overline{g}_k]$ which is equivalent to $r^q \in \mathcal{R}^q_k$, given the definition of $r^q_k$ in \eqref{eq:res_def}, and with $\underline{g}_k,\overline{g}_k$ obtained from \eqref{eq:nonl_bounding} in Corollary \ref{cor:bounding}.
\end{proof}
By Proposition \ref{prop:res_membership}, if the residual signal of a particular mode $q$ is not within the given interval in \eqref{eq:mode_elimination} conditioned on this mode being true, then $q$ can be ruled out as incompatible.
\subsubsection{Global Fusion Observer}
Finally, combining the outputs of both components above, our proposed global fusion observer will provide
 mode, attack and state set-valued estimates, as well as attack policy abstractions, at each time step $k$ as:
\begin{align*}
\begin{array}{c}
\hat{\mathcal{Q}}_k\hspace{-.1cm}=\hspace{-.1cm}\{q \in \hat{\mathcal{Q}}_{k-1} \, \vline \,  r^q_k \in \mathcal{R}^q_k \},\\ \hat{\mathcal{X}}_k=\cup_{q \in \hat{\mathcal{Q}}_k} \mathcal{X}^q_{k}, \hat{\mathcal{D}}_{k}=\cup_{q \in \hat{\mathcal{Q}}_k} \mathcal{D}^q_{k},
\\ \overline{\mu}_k(\cdot) = \max_{q \in \hat{\mathcal{Q}}} \overline{\mu}^q_k(\cdot), \ \underline{\mu}_k(\cdot) = \min_{q \in \hat{\mathcal{Q}}} \underline{\mu}^q_k(\cdot) .
\end{array}
\end{align*} 

\noindent The 
simultaneous mode, state and attack policy (SMSP) estimation approach is summarized 
 in Algorithm \ref{algorithm1}.
 \begin{algorithm}[!t] \small
\caption{Simultaneous Mode, State and Attack Policy (SMSP) Estimation}\label{algorithm1}
\begin{algorithmic}[1]
  \State $\hat{\mathcal{Q}}_0=\mathcal{Q}$;
  \For {$k =1$ to $N$}
  \For {$q \in \hat{\mathcal{Q}}_{k-1}$}
  \LineComment{Mode-Matched State and Attack Policy Set-Valued Estimates}
 \Statex \hspace{0.4cm} Compute $\overline{x}^q_k,\underline{x}^q_k,\overline{d}^q_k,\underline{d}^q_k$ through \eqref{eq:propagation}--\eqref{eq:ml};
 \LineComment{Mode Observer via Elimination}
 \Statex \hspace{0.4cm} $\hat{\mathcal{Q}}_k=\hat{\mathcal{Q}}_{k-1}$;
 \Statex \hspace{0.4cm} Compute $r^q_k$ via Definition \ref{defn:computedresidual};
  \If {\eqref{eq:mode_elimination} holds} $\hat{\mathcal{Q}}_k=\hat{\mathcal{Q}}_{k} \backslash \{q\}$;
 \EndIf
 \EndFor
 \LineComment{State and Input Estimates}
 \State $\hat{\mathcal{X}}_k=\cup_{q \in \hat{\mathcal{Q}}_k} \hat{\mathcal{X}}^q_k$; \ $\hat{\mathcal{D}}_k=\cup_{q \in \hat{\mathcal{Q}}_k} \hat{\mathcal{D}}^q_k$;
 \LineComment{Attack Policy Abstraction}
  \State $\overline{\mu}_k(\cdot) = \max_{q \in \hat{\mathcal{Q}}} \overline{\mu}^q_k(\cdot)$; \ $\underline{\mu}_k(\cdot) = \min_{q \in \hat{\mathcal{Q}}} \underline{\mu}^q_k(\cdot)$;
 \EndFor
\end{algorithmic}
\end{algorithm}
\vspace{-0.1cm}
\subsection{Properties of Mode-Matched Observers}
In this section, following a similar approach to our previous work \cite{khajenejad2021intervalACC}, we show that each of the mode-matched observers is correct (cf. Definition \ref{defn:framers}) and stable (cf. Definition \ref{defn:stability}) under some sufficient conditions. Moreover, the sequence of mode-matched interval widths is convergent to some computable steady state values.  
\begin{lem}[Correctness]\label{lem:correctness}
Consider System \eqref{eq:sys_desc} and suppose Assumptions \ref{ass:mixed_monotone}--\ref{ass:fixed_true_mode} hold. Then, for all mode $q \in \mathcal{Q}$, the dynamical system in \eqref{eq:propagation}--\eqref{eq:ml} constructs a correct mode-matched interval observer for System \eqref{eq:sys_desc}, \yo{conditioned on the mode being the true mode, i.e., $q =q^*$.} 
In other words, $\forall k  \in \mathbb{K} \triangleq \mathbb{N} \cup \{0\}, \underline{z}^q_k \leq z^q_k \leq \overline{z}^q_k$, where $z^q_k \triangleq [x^{\top} \ d^{q\top}]^\top$ and $[\overline{z}_k^{q\top} \underline{z}_k^{q\top}]^\top$ are the augmented vectors of state and unknown inputs in the dynamical systems in \eqref{eq:sys_desc} \yo{and the augmented estimates from \eqref{eq:iter_update}} 
at time $k \in {\mathbb{K}}$, 
respectively. 
\end{lem}
\begin{proof}
Using induction, the proof follows similar lines to the proof of \cite[Theorem 1]{khajenejad2021intervalACC}.
\end{proof}
Next, we address the stability of each mode-matched observer.
Note that similar to \cite{khajenejad2021intervalACC}, our goal is to obtain sufficient stability conditions that can be checked \emph{a priori} instead of for each time step $k$. On the other hand, for the implementation of the update step, we iteratively find new mode-matched \emph{local} parallel abstraction slopes $A^{q,g}_{i,k}$ by iteratively solving the LP \eqref{eq:abstraction} for $g^q$ on the intervals obtained in the previous iteration, $\mathcal{B}^{q,u}_{i,k}=[\underline{z}^{q,u}_{i-1,k},\overline{z}^{q,u}_{i-1,k}]$, to find \emph{local} framers $\overline{z}^{q,u}_{i,k},\underline{z}^{q,u}_{i,k}$ (cf. \eqref{eq:zupp}--\eqref{eq:alpha}), with additional 
constraints given in \eqref{eq:guarantee} in the optimization problems, which guarantees that the iteratively updated \emph{local} intervals obtained using the local abstraction slopes are inside the global interval, i.e.,
\begin{align*}
 \begin{array}{c}
  \underline{z}^{q,u}_k \leq \underline{z}^{q,u}_{0,k} \leq \dots \leq \underline{z}^{q,u}_{i,k} \leq \dots  \leq  \lim_{i \to \infty}\underline{z}^{q,u}_{i,k}  \triangleq \underline{z}^q_k,\\ \overline{z}^q_k \triangleq \lim_{i \to \infty}\overline{z}^{q,u}_{i,k}   \leq \dots\leq \overline{z}^{q,u}_{i,k} \leq \dots \leq \overline{z}^{q,u}_{0,k} \leq \overline{z}^{q,u}_k. 
  \end{array}
\end{align*}
With that in mind, we next show through the following proposition that the sequence of the widths of the interval-valued estimates are upper bounded by a difference equation,  i.e., a discrete-time dynamical system, for each mode.
\begin{prop}[Interval Widths Upper System]\label{prop:interval_width_system}
Consider System \eqref{eq:sys_desc} along with the observer in \eqref{eq:propagation}--\eqref{eq:ml} and suppose that all the assumptions in Lemma \ref{lem:correctness} hold and the decomposition function $f_d$ is constructed using \eqref{eq:decompconstruct}. Let us define the mode-matched width of the interval-valued estimate $[\underline{z}^q_k,\overline{z}^q_k]$, at time $k$, as $\Delta^{z^q}_k \triangleq \overline{z}^q_k-\underline{z}^q_k$. Then, for each mode $q \in \mathcal{Q}$, the following inequality holds for $k \in \mathbb{N}$: $\forall (D^q_1,D^q_2,D^q_3) \in 
\mathbb{D}_{n+p} \times \mathbb{D}_l \times \mathbb{D}_n $,
\begin{align} 
\Delta^{z^q}_k &\leq \mathcal{A}_q^g(D^q_1,D^q_2)\mathcal{A}^{f,h}_q(D^q_3)\Delta^{z^q}_{k-1} \label{eq:error_dynamics}\\
\nonumber &+\Delta^g_q(D^q_1,D^q_2)+\mathcal{A}_q^g(D^q_1,D^q_2) \Delta_q^{f,h}(D^q_3)+2 \kappa D^q_1\mathbf{r}^q,
\end{align}
where
\begin{align*}
 \mathcal{A}_q^g(D^q_1,D^q_2) &\triangleq D^q_1|A^{g\dagger}_q|D^q_2 |A_q^g|+(I-D^q_1),\\
  \mathcal{A}^{f,\mu}_q(D^q_3) &\triangleq \begin{bmatrix} (|A_q^f|+2(I-D^q_3)C^{f^q}_z)^\top & [|A_q^{\mu}| \ 0]^\top \end{bmatrix}^\top,\\
   \Delta^g_q(D^q_1,D^q_2) &\triangleq D^q_1|A_q^{g\dagger}|D^q_2(|W_q^g|\Delta v+\Delta^{g^q}_e),\\
\Delta^{f,h}_q(D^q_3) &\triangleq  (|W_q^f| \hspace{-.1cm}+ \hspace{-.1cm} 2(I-D^q_3)C^{f^q}_w) \Delta w \hspace{-.1cm}+ \hspace{-.1cm} \Delta^{f^q}_e,
 \end{align*}
while $\mathbf{r}^q \triangleq \textstyle{\mathrm{rowsupp}}(I-A^{g\dagger}_qA^g_q)$, $C_q^f \triangleq \begin{bmatrix} C^{f^q}_z & C^{f^q}_u & C^{f^q}_w \end{bmatrix}$ 
from \eqref{eq:decompconstruct}, $\kappa$ is a very large positive real number (infinity) and $\Delta^{g^q}_e \triangleq \overline{e}^{f^q}-\underline{e}^{f^q} ,\Delta^{f^q}_e \triangleq \overline{e}^{g^q}-\underline{e}^{g^q}, \Delta v \triangleq \overline{v}-\underline{v},\Delta w \triangleq \overline{w}-\underline{w}$, $\{A^s_q \triangleq \mathbb{A}^s_{(1:n+p)}\}_{s \in \{f^q,g^q\}}, A^{\mu}_q \triangleq \mathbb{A}^{{\mu}^q}, W^f_q \triangleq \mathbb{A}^s_{(n+p+1:n+p+n_w)},W^g_q \triangleq \mathbb{A}^g_{(n+p+1:n+p+n_v)}$, \yo{with $\mathbb{A}^s$ and $\mathbb{A}^{{\mu}^q}$ obtained using Proposition \ref{prop:affine abstractions}.} 
\end{prop}
\begin{proof}
The proof is similar to 
the proof of \cite[Theorem 2]{khajenejad2021intervalACC}, with some minor modifications, by replacing the unknown mapping $h$ with the unknown policy $\mu$ and making all variables mode-dependent.
\end{proof}
Now, armed with the results in Proposition \ref{prop:interval_width_system}, we provide sufficient conditions for the stability of each of the mode-matched observers in the sense of Definition \ref{defn:stability}, in a similar manner to \cite[Theorem 2]{khajenejad2021intervalACC}, through the following lemma.
\begin{lem}[Stability] \label{lem:stability}
Consider the hidden mode switched system \eqref{eq:sys_desc} along with the mode-matched observer in \eqref{eq:propagation}--\eqref{eq:ml}. Suppose that all the assumptions in Proposition \ref{prop:interval_width_system} hold.  
Then, for each mode $q \in \mathcal{Q}$, the mode-matched observer in \eqref{eq:propagation}--\eqref{eq:ml} is stable in the sense of Definition \ref{defn:stability},
if there exist $ D^q_1 \in \mathbb{D}_{n+p},D^q_2 \in \mathbb{D}_l,D^q_3 \in \mathbb{D}_n$ that satisfy $D^q_{1,i,i}=0$ if $\mathbf{r}^q(i)=1$, i.e., if there exist
$(D^q_1,D^q_2,D^q_3) \in \mathbb{D}^* 
\triangleq \{(D_1,D_2,D_3)\in \mathbb{D}_{n+p} \times \mathbb{D}_l \times \mathbb{D}_n \, \vline\, D_{1,ii}\mathbf{r}(i)=0 \}$ 
such that
\begin{align}
   \mathcal{L}^*(D^q_1,D^q_2,D^q_3) \triangleq \| \mathcal{A}_q^g(D^q_1,D^q_2)\mathcal{A}_q^{f,\mu}(D^q_3)\|  \leq 1, \label{eq:stability}
\end{align}
with $\mathcal{A}_q^g(D^q_1,D^q_2)$ and $\mathcal{A}_q^{f,\mu}(D^q_3)$ defined in Proposition \ref{prop:interval_width_system}.
\end{lem}
\begin{proof}
Our goal is to show that our specific choices for $D^q_1,D^q_2,D^q_3$, make the right hand side of \eqref{eq:error_dynamics} finite {in finite} time. To do this, since $\kappa$ can be infinitely large, 
we choose $D^q_1 \in \mathbb{D}_{n+p}$ such that $D^q_1\mathbf{r}^q=0$, i.e., $D^q_{1,i,i}=0 \ \text{if} \ \mathbf{r}^q(i)=1,  i=1,\dots, n+p$. Then, by the \emph{Comparison Lemma} \cite{khalil2002nonlinear}, it suffices for uniform boundedness of $\{\Delta^{z^q}_k\}_{k=0}^{\infty}$ that the following system:
\begin{align} \label{eq:eq:error_dynamics2}
\Delta^{z^q}_k &= \mathcal{A}_q^g(D^q_1,D^q_2)\mathcal{A}_q^{f,\mu}(D^q_3)\Delta^{z^q}_{k-1}+\tilde{\Delta}_q(D^q_1,D^q_2), 
\end{align}
 is stable, where $\tilde{\Delta}_q(D^q_1,D^q_2) \triangleq \Delta^g_q(D^q_1,D^q_2)+\mathcal{A}_q^g(D^q_1,D^q_2) \Delta_q^{f,\mu}(D^q_3)$ is a bounded disturbance. This implies that the system \eqref{eq:eq:error_dynamics2} is stable (in the sense of uniform stability of the interval sequnces) if and only if the matrix $\mathcal{A}_q(D^q_1,D^q_2,D^q_3) \triangleq \mathcal{A}_q^g(D^q_1,D^q_2)\mathcal{A}_q^{f,\mu}(D^q_3)$ is (non-strictly) stable for at least one choice of $(D^q_1,D^q_2,D^q_3) $, and equivalently, \eqref{eq:stability} 
 should hold. 
\end{proof}
Finally, the mode-matched interval widths are upper bounded and convergent to {steady-state values}, as follows.
\begin{prop}[Upper Bounds of the Interval Widths and their Convergence]\label{prop:convergence}
Consider the system \eqref{eq:sys_desc} and the observer \eqref{eq:propagation}--\eqref{eq:ml} and suppose all the assumptions in Lemma \ref{lem:stability} hold. Then, for each mode $q \in \mathcal{Q}$, the sequence of $\{\Delta^{z^q}_k \triangleq \overline{z}^q_k-\underline{z}^q_k\}_{k=0}^{\infty}$ is uniformly upper bounded by a convergent sequence, as $\Delta^{z^q}_k \leq \overline{\mathcal{A}}_q^{k}\Delta^{z^q}_0+\sum_{j=0}^{k-1} \overline{\mathcal{A}}_q^j\overline{\Delta}_q \xrightarrow{k \to \infty} e^{\overline{\mathcal{A}}_q} \overline{\Delta}_q$, 
where 
$\overline{\mathcal{A}}_q =\mathcal{A}_q(D^{q\star}_1,D^{q\star}_2,D^{q\star}_3) \triangleq \mathcal{A}-q^g(D^{q\star}_1,D^{q\star}_2)\mathcal{A}_q^{f,\mu}(D^{q\star}_3), 
\overline{\Delta}_q =\Delta_q^g(D^{q\star} _1,D^{q\star}_2)+\mathcal{A}_q^g(D^{q\star}_1,D^{q\star}_2) \Delta_q^{f,\mu}(D^{q\star}_3)$, 
and $(D^{q\star}_1,D^{q\star}_2,D^{q\star}_3)$ is a solution of the following problem:

\phantom{a}
\begin{small}

\vspace*{-.7cm}
\begin{align*}
&\min\limits_{D_1,D_2,D_3}\hspace{-.1cm} \|e^{\mathcal{A}_q(D_1,D_2,D_3)}(\Delta^g_q(D _1,D_2)\hspace{-.05cm}+\hspace{-.05cm}\mathcal{A}_q^g(D_1,D_2) \Delta^{f,\mu}_q(D_3))\| \\
&\quad \text{s.t.}  (D_1,D_2,D_3) \hspace{-.05cm} \in  \hspace{-.05cm} \{ (D_1,D_2,D_3)  \hspace{-.05cm} \in \hspace{-.05cm} \mathbb{D}^* \, \vline \, \mathcal{L}_q^*(D_1,D_2,D_3)  < 1 \}. 
\end{align*}
\end{small}

\vspace{-0.4cm}
Consequently, the interval widths $\{ \| \Delta^{z^q}_k \|\}_{k=1}^{\infty}$ are uniformly upper bounded by a convergent sequence, i.e., 
$\|\Delta^{z^q}_k\| \leq \delta^{z^q}_k \hspace{-.1cm} \triangleq \hspace{-.1cm} \| \overline{\mathcal{A}}_q^{k}\Delta^{z^q}_0\hspace{-.1cm}+\hspace{-.1cm}\sum_{j=0}^{k-1} \overline{\mathcal{A}}_q^j\overline{\Delta}_q\| \hspace{-.1cm} \xrightarrow{k \to \infty} \hspace{-.1cm} \|e^{\overline{\mathcal{A}}_q} \overline{\Delta}_q\|$.
\end{prop}
\begin{proof}
The proof is straightforward by applying \cite[Lemma 1]{efimov2013interval}, computing \eqref{eq:error_dynamics} iteratively, using triangle inequality and the fact that by Theorem \ref{lem:stability}, $\mathcal{A}_q(D^{q\star}_1,D^{q\star}_2,D^{q\star}_3)$ is a stable matrix and 
$(D^{q\star}_1,D^{q\star}_2,D^{q\star}_3)$ is a solution of \eqref{eq:stability}. 
\end{proof}
\section{Mode-Detectability}
 In addition to the nice properties regarding the correctness, stability and convergence of the mode-matched interval estimates of states and inputs, as discussed in the previous section, we now provide some sufficient conditions for the system dynamics and attack policies, which guarantee that regardless of the observations, after some large enough time steps, 
 \emph{all} the false (i.e., not true) modes can be eliminated, when applying Algorithm \ref{algorithm1}. 
 To do so, first, we define the concept of mode-detectability 
 as well as some assumptions for deriving our sufficient conditions for mode-detectability. 
\begin{defn}[Mode-Detectability] \label{defn:strong_mode_detecatble}
System \eqref{eq:sys_desc} is called mode-detectable under Algorithm \ref{algorithm1}, if there exists a natural number $K \in \mathbb{N}$, such that for all time steps $k \geq K$, all false modes are eliminated.
\end{defn}

\begin{assumption}[Destabilizing Attack Policy] \label{assumption:destabilizing}
For all $q \in \mathcal{Q}$, the vector fields $ f^{q}(x,\mu^q(x),w)$ and $\mu^{q}(x)$ satisfy the following bounds on their Jacobians: $\forall j \in \{x,d\}, \forall (x,w) \in \mathcal{X} \times \mathcal{W}$, $J_j^{f^q}(x,\mu^q(x),w) \in [\underline{J}^{f^q}_j,\overline{J}^{f^q}_j]$ and $J^{\mu^q}(x) \in [\underline{J}^{\mu^q},\overline{J}^{\mu^q}]$, with known $\overline{J}^{f^q},\underline{J}^{f^q},\overline{J}^{\mu^q},\overline{J}^{\mu^q}$ a priori, where $J^{f^q}_x,J^{f^q}_d$ are Jacobians of $ f^{q}$, with respect to its first argument, $x$, and second argument $d$, respectively. Moreover, $J_q^{\text{\emph{m}}} \triangleq \frac{1}{2}(\underline{J}^{f^q}_x+\overline{J}^{f^q}_x+\underline{J}^{f^q,\mu^q}_d+\overline{J}^{f^q,\mu^q}_d)$ is  strictly Schur unstable\footnote{A strictly Schur unstable matrix is a square matrix that has  at least one eigenvalue with its real part being stricly outside the interval $[0,1]$.}, where $\underline{J}^{f^q,\mu^q}_d \triangleq \underline{J}^{f^q+}_d\underline{J}^{\mu^q+}-\overline{J}^{f^q+}_d\underline{J}^{\mu^q-}-\underline{J}^{f^q-}_d\overline{J}^{\mu^q+}+\overline{J}^{f^q-}_d\overline{J}^{\mu^q-}$ and $\overline{J}^{f^q,\mu^q}_d \triangleq \overline{J}^{f^q+}_d\overline{J}^{\mu^q+}-\underline{J}^{f^q+}_d\overline{J}^{\mu^q-}-\overline{J}^{f^q-}_d\underline{J}^{\mu^q+}+\underline{J}^{f^q-}_d\underline{J}^{\mu^q-}$.
\end{assumption}

\begin{cor}\label{cor:destabilizing}
Assumption \ref{assumption:destabilizing} implies that the attack policy $\mu(\cdot) \triangleq \mu^{q^*}(\cdot)$ destabilizes the system in \eqref{eq:sys_desc}, and hence, $x_k$, the true state trajectory of \eqref{eq:sys_desc} becomes unbounded. 
\end{cor}
\begin{proof}
Defining $f\triangleq f^{q^*}$, $\mu(x) \triangleq \mu^{q^*}(x)$ and $\tilde{f}(x,w) \triangleq f(x,\mu(x),w)$, as well as using \emph{chain rule}, we have $J^{\tilde{f}}_x=J_x^{f}(x,\mu(x),w)+J_d^f(x,\mu(x),w)J^{\mu}(x)$. Combining this and Assumption \ref{assumption:destabilizing}, as well as applying Proposition \ref{prop:uncertain_bounding}, returns $J^{\tilde{f}}_x \in [\underline{J}_x^{f}+\underline{J}_d^{f,\mu},\overline{J}_x^{f}+\overline{J}_d^{f,\mu}]$. Now, note that since $J_{q^*}^{\text{m}}$ is strictly Schur unstable by Assumption \ref{assumption:destabilizing}, then the interval matrix $J^{\tilde{f}}_x $ is strictly Schur unstable by \cite[Lemma 2b]{wang1994necessary}, and hence, the linearized form of the system in \eqref{eq:sys_desc} is strictly Schur unstable. Consequently, the nonlinear system in \eqref{eq:sys_desc} is unstable by the Chetaev instability theorem \cite{chetaev1962stability}, i.e., the attack policy $\mu$ is a destabilizing policy.
\end{proof}
Now, we are ready to state our main result on mode-detectability, through the following theorem.
\begin{thm}[Sufficient Conditions for Mode-Detectability] \label{thm:strong_mode_detect}
Suppose Assumption \ref{assumption:destabilizing} and all the assumptions in Lemma \ref{lem:stability} hold for all $q \in \mathcal{Q}$. Then, using Algorithm \ref{algorithm1}, System \eqref{eq:sys_desc} is mode-detectable in the sense of Definition \ref{defn:strong_mode_detecatble}.
\end{thm}
\begin{proof}
We need to show that there exists $K \in \mathbb{N}$, such that \eqref{eq:mode_elimination} holds for all $k \geq K, \forall q \neq q^* \in \mathcal{Q}$, where $q^*$ is the true mode. Given the definition of the residual signal in \eqref{eq:res_def} and since $q^*$ is unknown, a sufficient condition for \eqref{eq:mode_elimination} to hold is that $\forall q_1 \neq q_2 \in \mathcal{Q},\exists K \in \mathbb{N}, \forall k \geq K, g^{q_2}(\xi^{q_2}_k) \notin [\underline{g}^{q_1}_k,\overline{g}^{q_1}_k]$, where $\forall q \in \mathcal{Q}, \xi^{q}_k \triangleq [x^\top_k \  d^{q\top}_k \ v^{q\top}_k]$. Equivalently, there should exist a dimension $i \in \mathbb{N}_l$, such that $g_i^{q_2}(\xi^{q_2}_k) < \underline{g}^{q_1}_{i,k}$ or $g^{q_2}_i(\xi^{q_2}_k) > \overline{g}^{q_1}_{i,k}$. Since $q_1 \neq q_2$ can be any two arbitrary modes, then without loss of generality, we only consider the former inequality, for which to hold, a sufficient condition is that $\overline{g}^{q_2}_{i,k} < \underline{g}^{q_1}_{i,k}$, which by defining $\overline{\Delta}^{q}_{g_i} \triangleq \overline{g}^{q}_{i,k}-g_i^{q}(\xi^{q}_k) \geq 0$ and $\underline{\Delta}^{q}_{g_i} \triangleq g_i^{q}(\xi^{q}_k)-\underline{g}^{q}_{i,k} \geq 0$, is equivalent to $g_i^{q_2}(\xi^{q_2}_k)+\overline{\Delta}^{q_2}_{g_i}  < g_i^{q_1}(\xi^{q_1}_k)-\underline{\Delta}^{q_1}_{g_i}$, that can be rewritten as:
\begin{align}\label{eq:mode_det_equivalent}
 \overline{\Delta}^{q_2}_{g_i} +\underline{\Delta}^{q_1}_{g_i} \leq g_i^{q_1}(\xi^{q_1}_k)-g_i^{q_2}(\xi^{q_2}_k).
 \end{align}  
 Now, note that the left hand side of \eqref{eq:mode_det_equivalent} can be verified to be bounded as follows: $0 \leq  \overline{\Delta}^{q_2}_{g_i} +\underline{\Delta}^{q_1}_{g_i} \leq \Delta^{q_2}_{g_i}+\Delta^{q_1}_{g_i}$, where $\forall q \in \mathcal{Q}, \Delta^q_{g_i} \triangleq \overline{g}^q_{i,k}-\underline{g}^q_{i,k}$ is bounded by the the Lipschitz-like property of the decomposition functions (cf. \cite[Lemma 2]{khajenejad2021intervalACC}), as well as the stability of each of the mode-matched observers (cf. Lemma \ref{lem:stability}). Now that the left hand side of \eqref{eq:mode_det_equivalent} is proven to be bounded, if we show that the right hand side grows unboundedly, then \eqref{eq:mode_det_equivalent} must always hold after some sufficiently large time step $K \in \mathbb{N}$. To do so, we consider some $x_0 \in \mathcal{X}_0$ and apply the mean value theorem on both $g^{q_1}_i$ and $g^{q_2}_i$ to obtain $g_i^{q_1}(\xi^{q_1}_k)-g_i^{q_2}(\xi^{q_2}_k)=(J^{q_2}_{g,i}(\tilde{\xi}_{k}^2)-J^{q_1}_{g,i}(\tilde{\xi}_{k}^1))(x_k-x_0)$, where $\tilde{\xi}^2_k,\tilde{\xi}^1_k \in \mathcal{X} \times \mathbb{R}^p \times \mathcal{V}$ and for all $q \in \mathcal{Q}$, $J^q_{g}(\xi)$ and $J^q_{g,i}(\xi)$ denote the Jacobian matrix of $g^q$ and its $i$'th row, evaluated at $\xi$. Finally, the right hand side of the above equality eventually becomes unbounded, since the Jacobian matrix of $J^q$ is bounded for all $q \in \mathcal{Q}$ by Lipschitz continuity of $g^{q}$ (cf. Assumption \ref{ass:mixed_monotone}), and $x_k$, i.e., the true state trajectory becomes unbounded by Assumption \ref{assumption:destabilizing} and Corollary \ref{cor:destabilizing}. Hence, the left hand side of the above equality also becomes eventually unbounded, which returns the desired result. 
\end{proof}
\vspace{-0.2cm}
\section{Simulation Example} \label{sec:examples}
\begin{figure}[t]
\centering
\includegraphics[width =0.45\textwidth]{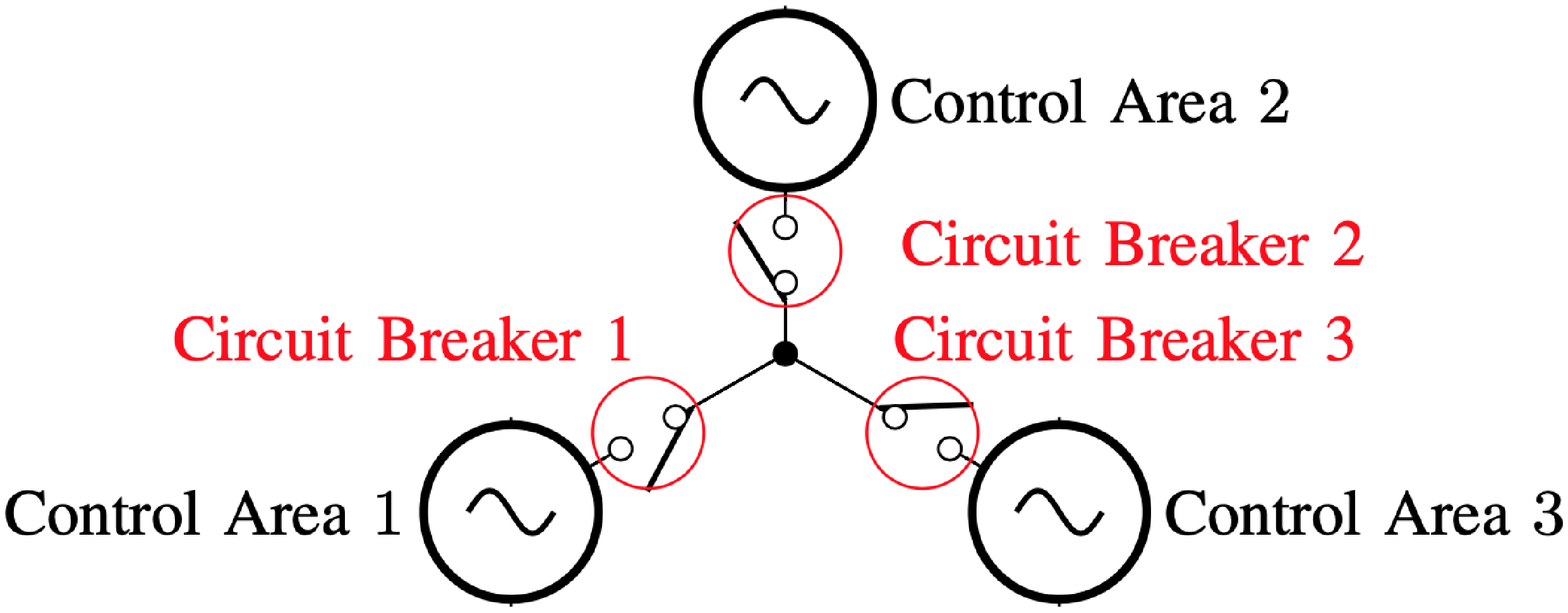}
\caption{Example of a three-area power station in a radial
topology (corresponding to node/bus attack).}
\label{fig:3-area}
\end{figure}

In this section, we illustrate \mo{the effectiveness} our proposed \mo{observer} using a power network with multiple control areas. Specifically, we consider a 3-area system as shown in Figure \ref{fig:3-area} where each control area consists of generator and load buses. \mo{In addition, there are} transmission lines between areas.
The nonlinear model of bus $i$ is adopted from \cite{kim2016attack}:
\begin{align}
 \nonumber   \dot{\theta}_i(t) &= f_i(t) + w_{1, i}(t), \\
 \nonumber  \dot{f}_i(t) &\hspace{-.1cm}= \hspace{-.1cm}-\frac{D_i f_i(t) \hspace{-.1cm}+\hspace{-.1cm} \Sigma P_{il}(t) \hspace{-.1cm}-\hspace{-.1cm} (P_{M_i}(t) \hspace{-.1cm}+\hspace{-.1cm} d_{ i}(t))\hspace{-.1cm}+\hspace{-.1cm}P_{L_i}(t)}{m_i}\hspace{-.1cm}+ \hspace{-.1cm}w_{2, i}(t),
\end{align}
with the output model: $y_{i, k} = [\theta_{i,k}, f_{i, k}]^\top + [0, 1]^\top d_{i} + v_{i, k}$,
where $\theta_i$ is \mo{phase angle}, $f_i$ is \mo{the angular} frequency, $P_{M_i}(t)$ \mo{denotes the} mechanical power 
(the control input) and $P_{L_i}(t)$ is a \mo{known} power demand. In our simulations, both $P_{M_i}(t)$ and $P_{L_i}(t)$ are set to be identically zero and the process noise $w_{i}(t)$ and measurement noise $v_{i}(t)$ are bounded by $\begin{bmatrix} -0.1 & 0.1 \\ -0.1 &0.1 \end{bmatrix}$ and $\begin{bmatrix} -0.1 & 0.1 \\ -0.1 &0.1 \end{bmatrix}$, respectively. 
When the circuit breakers are not engaged (or attacked), 
the power flow $P_{il}$ between areas $i$ and $l$ is as follows: 
\begin{align*}
 P_{il}(t) = -P_{li}(t) = t_{il}\sin(\theta_i(t) - \theta_j(t)).    
\end{align*}
A malicious agent is assumed to have access to circuit
breakers that control the tie-lines, and is thus able to sever the connection between control areas. Two types of attack are considered based on the topology of the tie-line interconnection graph: (1) a node/vertex/bus attack (disconnection of a control area from all others); or (2) a link/edge/line attack (disabling of a specific tie-line between two control areas), i.e., the power flow across the tie lines is altered, if (1) there is an
attack on \yo{control area} $i$ (node/bus attack): $P_{il}(t) = -P_{li}(t) = 0, \ \forall l \neq i$; or (2) if there exists an
attack on circuit breaker $(i,l)$ (link/line attack):
    $P_{il}(t) = -P_{li}(t) = 0$. 

For the radial tie-line interconnection topology in Figure \ref{fig:3-area}, the circuit breaker attacks result in $Q=5$ possible modes of operation: all switches are safe ($q=1$), only circuit breaker $i$ is attacked ($q=i+1, \ i = 1, 2, 3$) and two or more circuit breakers are attacked ($q=5$).
Further, we denote the value \mo{of the variables} at sampling time $t_k$ by adding subscript $k$, e.g., $f_i(t_k) = f_{i, k}$ and apply the Euler method to discretize the system: 
$\theta_{i, k+1} = \theta_{i, k} + \dot{\theta}_{i, k}dt, \
f_{i, k+1} = f_{i, k} + \dot{f}_{i, k} dt$,
where \mo{the sampling time} $dt$ 
is $0.01s$ in our example. 
Moreover, we choose 
$d_{i}(t) = \theta_i(t) \sin(\theta_i(t))$  as to-be-learned unknown nonlinear attack policy and assume that we have \yo{$400$} initial data points for each unknown attack policy $d_i$.

\begin{figure}[t]
\centering
\includegraphics[scale=0.26,trim=2mm 0mm 0mm 0mm,clip]{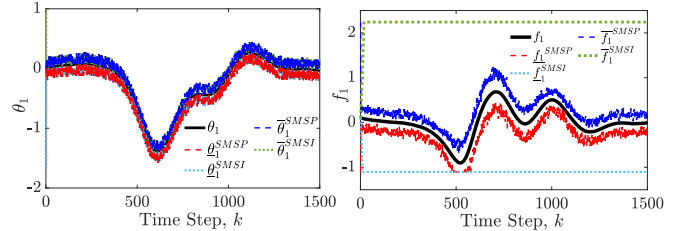} 
\caption{\mo{The true values of the states: $\theta_1,f_1$, and their upper and lower framers returned by the SMSP approach: $\overline{\theta}^{SMSP}_1,\underline{\theta}^{SMSP}_1,\overline{f}^{SMSP}_1,\underline{f}^{SMSP}_1$, as well as the SMSI approach: $\overline{\theta}^{SMSI}_1,\underline{\theta}^{SMSI}_1,\overline{f}^{SMSI}_1,\underline{f}^{SMSI}_1$}.}
\label{fig:states}
\end{figure}

\begin{figure}[t]
\centering
\includegraphics[scale=0.26,trim=2mm 0mm 0mm 0mm,clip]{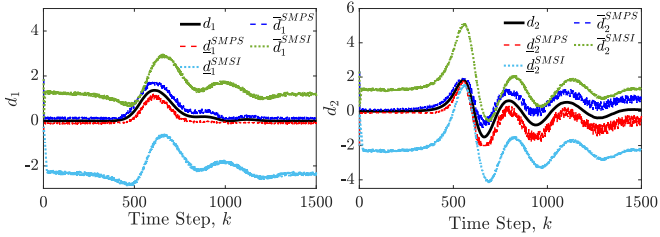} 
\caption{\mo{The true values of the attacks: $d_1,d_2$, and their upper and lower framers returned by the SMSP approach: $\overline{d}^{SMSP}_1,\underline{d}_1^{SMSP},\overline{d}^{SMSP}_2,\underline{d}^{SMSP}_2$, as well as the SMSI approach: $\overline{d}^{SMSI}_1,\underline{d}^{SMSI}_1,\overline{d}^{SMSI}_2,\underline{d}^{SMSI}_2$}.}
\label{fig:attacks}
\end{figure}

\mo{Due to space limitations}, we only show the results for the case when the true operation mode is assumed to be $q* = 1$ and provide figures for selected states and attack signals in Figure \ref{fig:states} and \ref{fig:attacks}. Moreover, we compare our result\mo{s} with \mo{our previously developed simultaneous mode, state and unknown input (SMSI) observer} in \cite{khajenejad2019simultaneousmode}, \mo{where no unknown policy/feedback law was assumed to govern the attack signals, and hence, no learning step were included in the proposed observer design}. \mo{As can be observed from Figures \ref{fig:states} and \ref{fig:attacks}}, \mo{the SMSP observer (proposed in this paper) returns tighter interval estimates than SMSI for} both states and attack signals, when compared to SMSI. \mo{It is also worth mentioning that} all the state \mo{interval widths} converge to steady state \mo{values} by using SMSP with ``learned" model 
\mo{for the attack policy}, while \mo{the interval widths for} some states do not converge \mo{when applying SMSI (not depicted for brevity), which highlights the effectiveness of the learning step}. 

\begin{figure}[t]
\centering
\includegraphics[width =0.235\textwidth,trim=5mm 0mm 15mm 0mm]{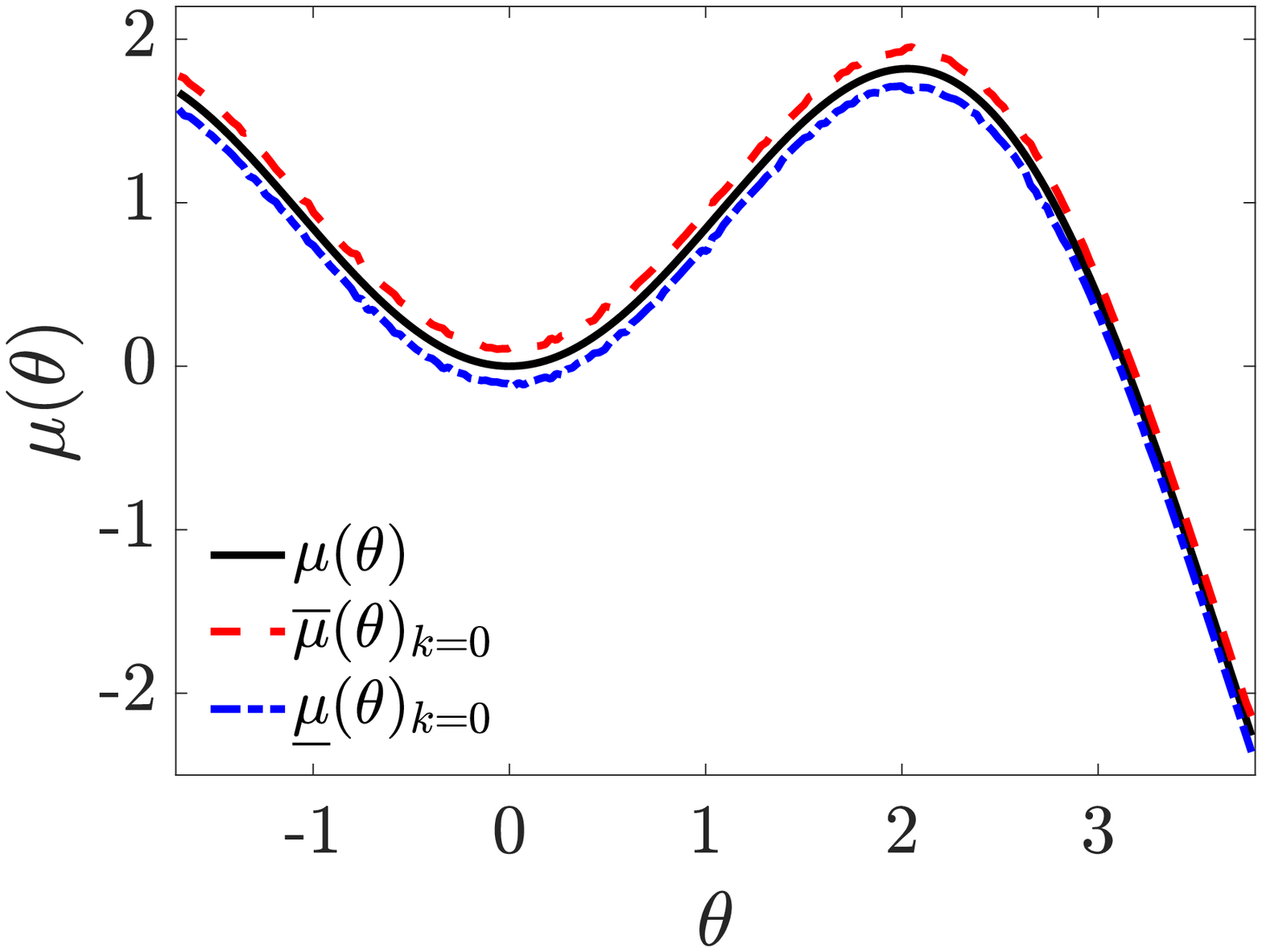} \includegraphics[width =0.235\textwidth,trim=5mm 0mm 15mm 0mm]{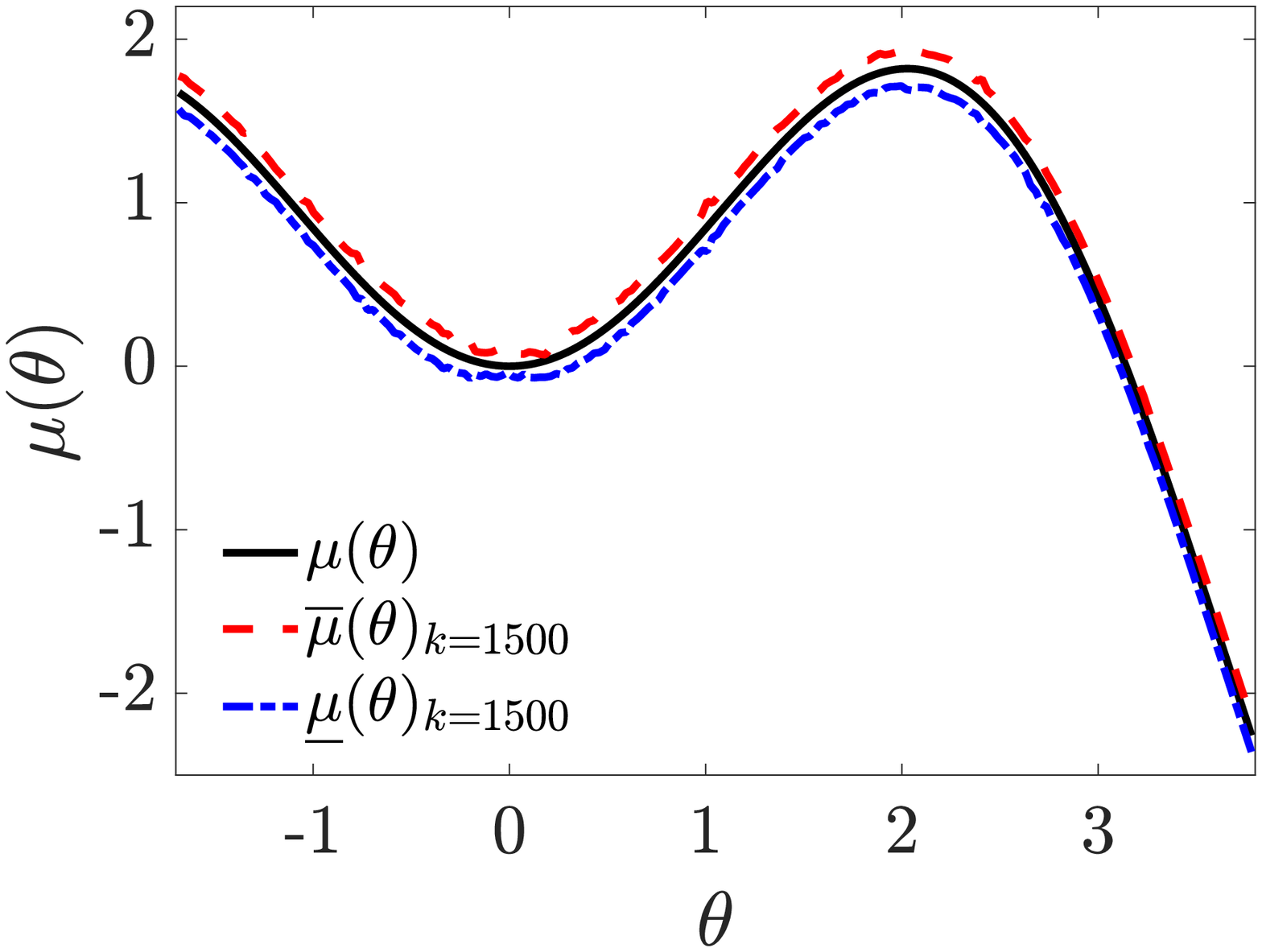} 
\caption{\mo{The true attack policy $\mu(\cdot)$, and its learned abstraction model $\{\underline{\mu}(\cdot),\overline{\mu}(\cdot)\}$} at time steps $k=0$ and $k=1500$.}
\label{fig:learn}
\end{figure}

\begin{figure}[t]
\centering
\includegraphics[width =0.525\textwidth,trim=20mm 110mm 0mm 0mm,clip]{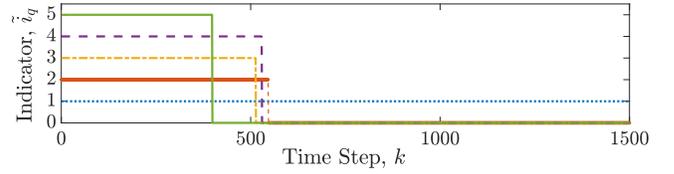} 
\caption{\mo{Mode estimates with indicators $\tilde{i}_q=qi_q$, where $i_q=0$ if mode $q$ is eliminated and $i_q=1$ otherwise}.}
\label{fig:mode}
\end{figure}

\begin{figure}[t]
\centering
\includegraphics[scale=0.325,trim=10mm 20mm 5mm 5mm,clip]{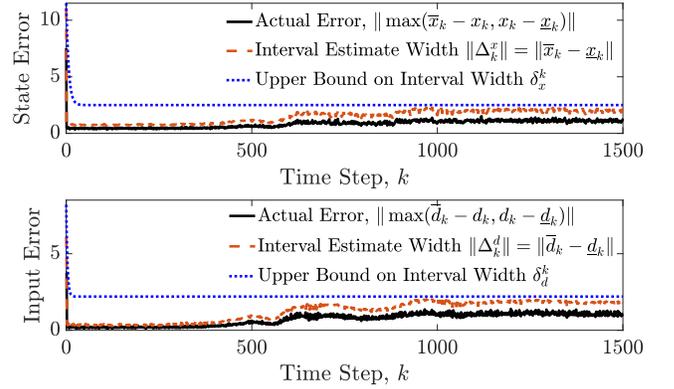} 
\caption{\mo{State and input} estimation error sequences.}
\label{fig:error}
\end{figure}

Moreover, we compare the upper and lower learned model abstractions for $k=0$ and $k=1500$ in Figure \ref{fig:learn}, which showed tighter over-approximations with increasing number of data points. 
\mo{Further, as can be seen in} Figure \ref{fig:mode}, \mo{all modes, except the true mode $q^*=1$, are eliminated} within $1500$ time steps. \mo{Finally}, the actual state and input estimation error sequence, as shown in Figure \ref{fig:error}, is upper bounded by the interval widths and converges to steady-state values. 

\section{Conclusion} \label{sec:conclusion}
  This paper addresses the problem of designing interval observers for hidden mode switched nonlinear systems with bounded noise signals, that are compromised by false data injection and switching attacks. 
  An interval observer with three constituents was proposed: i) a bank of mode-matched observers, where each of them simultaneously outputs the corresponding mode-matched state, mode and unknown input/attack estimates, as well as computes upper and lower abstractions/over-approximations for the attack policies, ii) a mode estimator that rules out the incorrect modes based on a residual-based set-membership criterion, and iii) a global fusion observer that returns the union of compatible state and attack estimates, as well as learned abstractions of the attack policy/state feedback law. 
Moreover, sufficient conditions for mode-detectability, i.e., for guaranteeing that all false modes will be eliminated after sufficiently large finite time steps, were provided. Finally, the effectiveness and performance of our proposed approach was demonstrated using a 3-area power network. 
\normalsize


\bibliographystyle{unsrturl}

\bibliography{biblio}
\end{document}